\documentclass[aps,showpacs,twocolumn,twoside,prc]{revtex4-1}
\usepackage{graphicx}% Include figure files
\usepackage{epstopdf}
\usepackage{dcolumn}% Align table columns on decimal point
\usepackage[ulem=normalem]{changes}
\usepackage{amsmath}
\usepackage{amssymb}

\newcommand{\ba}{\begin{eqnarray}}
\newcommand{\ea}{\end{eqnarray}}
\newcommand{\bmath}{\begin{mathletters}}
\newcommand{\emath}{\end{mathletters}}
\newcommand{\ban}{\begin{eqnarray*}}
\newcommand{\ean}{\end{eqnarray*}}

\newcommand{\bsub}{\begin{subequations}}
\newcommand{\esub}{\end{subequations}}

\def\ket#1{|#1\rangle}
	
\def\bsu3{\overline{{\rm SU(3)}}}
\def\blam{\bar{\lambda}}
\def\bmu{\bar{\mu}}
\def\bK{\bar{K}}
\def\b0{\beta_0}
\def\g0{\gamma_0}
\def\bs{\beta_*}
\def\bss{\beta_{**}}

\begin{document}
\title{
Algebraic benchmark for prolate-oblate coexistence in nuclei}
\author{A.~Leviatan}
\author{D.~Shapira}

\affiliation{Racah Institute of Physics, The Hebrew University, 
Jerusalem 91904, Israel}
\date{\today}

\begin{abstract}
We present a symmetry-based approach for prolate-oblate 
and spherical-prolate-oblate shape coexistence, 
in the framework of the interacting boson model of nuclei. 
The proposed Hamiltonian conserves the SU(3) and $\overline{\rm SU(3)}$
symmetry for the prolate and oblate ground bands and the U(5) symmetry 
for selected spherical states. Analytic expressions for quadrupole 
moments and $E2$ rates involving these states are derived and 
isomeric states are identified by means of selection rules.
\end{abstract}
\maketitle

A prominent feature in atomic nuclei, exemplifying a quantal mesoscopic 
system, is their ability to accommodate distinct shapes in 
their low-lying spectrum. 
Such shape coexistence in the same nucleus is known to occur widely 
across the nuclear chart~\cite{Heyde11,Focus16}.
The increased availability of rare isotope beams and advancement in 
high-resolution spectroscopy, open new capabilities to investigate such 
phenomena in nuclei far from stability~\cite{Jenkins14}. Notable 
empirical examples include the coexistence of prolate and oblate shapes 
in the neutron-deficient Kr~\cite{Clement07}, 
Se~\cite{Ljun08} and Hg~\cite{Bree14} isotopes, and the triple coexistence 
of spherical, prolate and oblate shapes in $^{186}$Pb~\cite{Andreyev00}. 
A detailed microscopic interpretation of nuclear shape-coexistence
is a formidable task. In a shell model description of nuclei 
near shell-closure, it is attributed to the occurrence of 
multi-particle multi-hole intruder excitations. 
For medium-heavy nuclei, this necessitates drastic truncations of 
large model spaces, {\it e.g.}, by Monte Carlo 
sampling~\cite{MCSM99,Tsunuda14}
or by a bosonic approximation of nucleon 
pairs~\cite{Foisson03,Frank04,Morales08,ramos14,
robeldo12,nomura13,Nomura16}. 
In a mean-field approach, based on energy density functionals, 
the coexisting shapes are associated with different minima 
of an energy surface calculated self-consistently. 
A detailed comparison with spectroscopic observables 
requires beyond mean-field methods, 
including restoration of broken symmetries and configuration mixing of 
angular-momentum and particle-number projected 
states~\cite{Bender13,vretenar16}. 
Such extensions present a major computational effort and often 
require simplifying assumptions such as axial symmetry 
and/or a mapping to collective model 
Hamiltonians~\cite{robeldo12,nomura13,
Nomura16,Bender13,vretenar16,EDF-IBM11}. 

A recent global mean-field calculation of nuclear shape isomers, 
identified experimentally accessible regions of nuclei 
with multiple minima in their potential-energy 
surface~\cite{moller09,moller12}.  
Such heavy-mass nuclei awaiting exploration, 
are beyond the reach of realistic large-scale shell model calculations. 
With that in mind, we present a simple 
alternative to describe coexistence of prolate 
and oblate shapes in large deformed nuclei, away from shell-closure, in the 
framework of the interacting boson model (IBM)~\cite{ibm}. 
The proposed approach emphasizes the role of remaining underlying 
symmetries which provide physical insight and make the problem tractable.

The IBM has been widely used to describe quadrupole collective states 
in nuclei in terms of $N$ 
monopole ($s^\dag$) and quadrupole ($d^\dag$) bosons,
representing valence nucleon pairs.
The model has U(6) as a spectrum generating algebra and exhibits 
several dynamical symmetries (DSs) 
associated with chains of nested subalgebras. 
These solvable limits admit analytic solutions, 
with closed expressions for observables, 
quantum numbers and definite selection rules. 
The DS chains with leading subalgebras: 
U(5), SU(3), ${\rm\overline{SU(3)}}$ and SO(6) 
correspond to known paradigms of nuclear collective 
structure: spherical vibrator, prolate-, oblate- and $\gamma$-soft deformed 
rotors, respectively. 
This identification is consistent with the geometric visualization 
of the model, obtained by an energy surface, $E_{N}(\beta,\gamma)$, 
defined by the expectation value of the Hamiltonian in the coherent 
(intrinsic) state~\cite{gino80,diep80},
\ba
\vert\beta,\gamma ; N \rangle =
(N!)^{-1/2}(b^{\dagger}_{c})^N\,\vert 0\,\rangle ~, 
\label{int-state}
\ea
where $b^{\dagger}_{c}\propto 
\beta\cos\gamma
d^{\dagger}_{0} + \beta\sin{\gamma}
( d^{\dagger}_{2} + d^{\dagger}_{-2})/\sqrt{2} + s^{\dagger}$.
Here $(\beta,\gamma)$ are
quadrupole shape parameters whose values 
at the global minimum of $E_{N}(\beta,\gamma)$ define the equilibrium
shape for a given Hamiltonian. 
The shape can be spherical $(\beta \!=\!0)$ or 
deformed $(\beta >0)$ with $\gamma \!=\!0$ (prolate), 
$\gamma \!=\!\pi/3$ (oblate), 
$0 \!<\! \gamma \!<\! \pi/3$ (triaxial) or $\gamma$-independent.
The standard DS Hamiltonians support a single minimum in their 
energy surface, hence serve as benchmarks for the dynamics of a single 
quadrupole shape.
In the present Rapid Communication, we propose a novel algebraic benchmark 
for describing the coexistence of prolate-oblate (P-O) shapes with equal 
$\beta$-deformations and triple coexistence of 
spherical-prolate-oblate (S-P-O) shapes. We focus on the dynamics in the 
vicinity of the critical point where the corresponding multiple minima 
are near-degenerate.

The DS limits appropriate to prolate and oblate shapes correspond, 
respectively, to the chains~\cite{ibm}
\bsub
\ba
{\rm U(6)\supset SU(3)\supset SO(3)} &&\;\;\quad
\ket{N,\, (\lambda,\mu),\,K,\, L} ~,\quad 
\label{SU3}
\\
{\rm U(6)\supset \bsu3\supset SO(3)} &&\;\;\quad
\ket{N,\, (\blam,\bmu),\,\bar{K},\, L} ~.\quad
\label{SU3bar}
\ea
\label{chains}
\esub
The indicated basis states are specified by quantum~numbers which are the 
labels of irreducible representations (irreps) of the algebras in each chain.
For a given $N$, the allowed SU(3) [$\,\bsu3\,$] irreps are 
$(\lambda,\mu)\!=\!(2N \!-\! 4k \!-\! 6m,2k)$ 
[$(\blam,\bmu)\!=\!(2k,2N\!-\!4k\!-\!6m)$] 
with $k,m$, non-negative integers. 
The multiplicity label $K$ ($\bK$) corresponds geometrically to the
projection of the angular momentum ($L$) on the symmetry axis. 
The basis states are eigenstates of the Casimir operator 
$\hat{C}_{2}[SU(3)]$ or $\hat{C}_{2}[\bsu3]$, 
where $\hat{C}_{2}[SU(3)] \!=\! 2Q^{(2)}\cdot Q^{(2)} \!+\! 
{\textstyle\frac{3}{4}}L^{(1)}\cdot L^{(1)}$, 
$Q^{(2)} \!=\! d^{\dagger}s \!+\! s^{\dagger}\tilde{d} 
\!-\!\frac{1}{2}\sqrt{7} (d^{\dagger}\tilde{d})^{(2)}$, 
$L^{(1)} \!=\! \sqrt{10} (d^{\dagger}\tilde{d})^{(1)}$, 
$\tilde{d}_{\mu} \!=\! (-1)^{\mu}d_{-\mu}$
and $\hat{C}_{2}[\bsu3]$ is obtained by replacing $Q^{(2)}$ by 
$\bar{Q}^{(2)} \!=\! d^{\dagger}s \!+\! s^{\dagger}\tilde{d} 
\!+\!\frac{1}{2}\sqrt{7} (d^{\dagger}\tilde{d})^{(2)}$. 
The generators of SU(3) and $\bsu3$, $Q^{(2)}$ and $\bar{Q}^{(2)}$, 
and corresponding basis states, are related 
by a change of phase $(s^{\dag},s)\rightarrow (-s^{\dag},-s)$, 
induced by the operator ${\cal R}_s=\exp(i\pi\hat{n}_s)$, 
with $\hat{n}_s=s^{\dag}s$. 
The DS Hamiltonian involves a linear combination of the  Casimir operators 
in a given chain. The spectrum resembles that of an axially-deformed 
rotovibrator composed of SU(3) [or $\bsu3$] multiplets forming 
rotational bands, with $L(L+1)$-splitting generated by 
$\hat{C}_{2}[{\rm SO(3)}] \!=\! L^{(1)}\cdot L^{(1)}$. 
In the SU(3) [or $\bsu3$] DS limit, the lowest irrep $(2N,0)$ [or $(0,2N)$] 
contains the ground band $g(K\!=\!0)$ [or $g(\bK\!=\!0)$] 
of a prolate (oblate) deformed nucleus. 
The first excited irrep $(2N\!-\!4,2)$ [or $(2,2N\!-\!4)$] contains 
both the $\beta(K\!=\!0)$ and $\gamma(K\!=\!2)$ 
[or $\beta(\bK\!=\!0)$ and $\gamma(\bK\!=\!2)$] bands. 
Henceforth, we denote such prolate and oblate bands by 
$(g_1,\beta_1,\gamma_1)$ and ($g_2,\beta_2,\gamma_2$), respectively. 
Since ${\cal R}_sQ^{(2)}{\cal R}_s^{-1} \!=\! -\bar{Q}^{(2)}$, 
the SU(3) and $\bsu3$ DS spectra are identical and 
the quadrupole moments of corresponding states differ in sign.

The phase transition between prolate and oblate shapes has been 
previously studied by varying a control parameter 
in the IBM Hamiltonian~\cite{jolie01,jolie03}. 
The latter, however, consisted of one- and two-body terms 
hence could not accommodate simultaneously two deformed minima. 
A solvable albeit schematic description of asymmetric P-O shapes 
was analyzed in~\cite{draayer12}, with higher-order 
SU(3)-invariant IBM interactions. 
P-O coexistence was considered within the IBM with 
configuration mixing, by using different Hamiltonians for the normal 
and intruder configurations and a number-non-conserving mixing 
term~\cite{Foisson03,Frank04,Morales08}. 
In the present work, we adapt a different strategy. We construct 
a single number-conserving Hamiltonian 
which retains the virtues of SU(3) and $\bsu3$ DSs for the 
prolate and oblate ground bands and the U(5) DS for selected 
spherical states. 
The construction is based on 
an intrinsic-collective resolution of the 
Hamiltonian~\cite{kirlev85,Leviatan87,levkir90}, 
a procedure used formerly in the study of spherical-deformed shape 
coexistence~~\cite{Leviatan06,Leviatan07,Macek14}. 
\begin{figure}[t]
  \centering
\includegraphics[width=8.6cm]{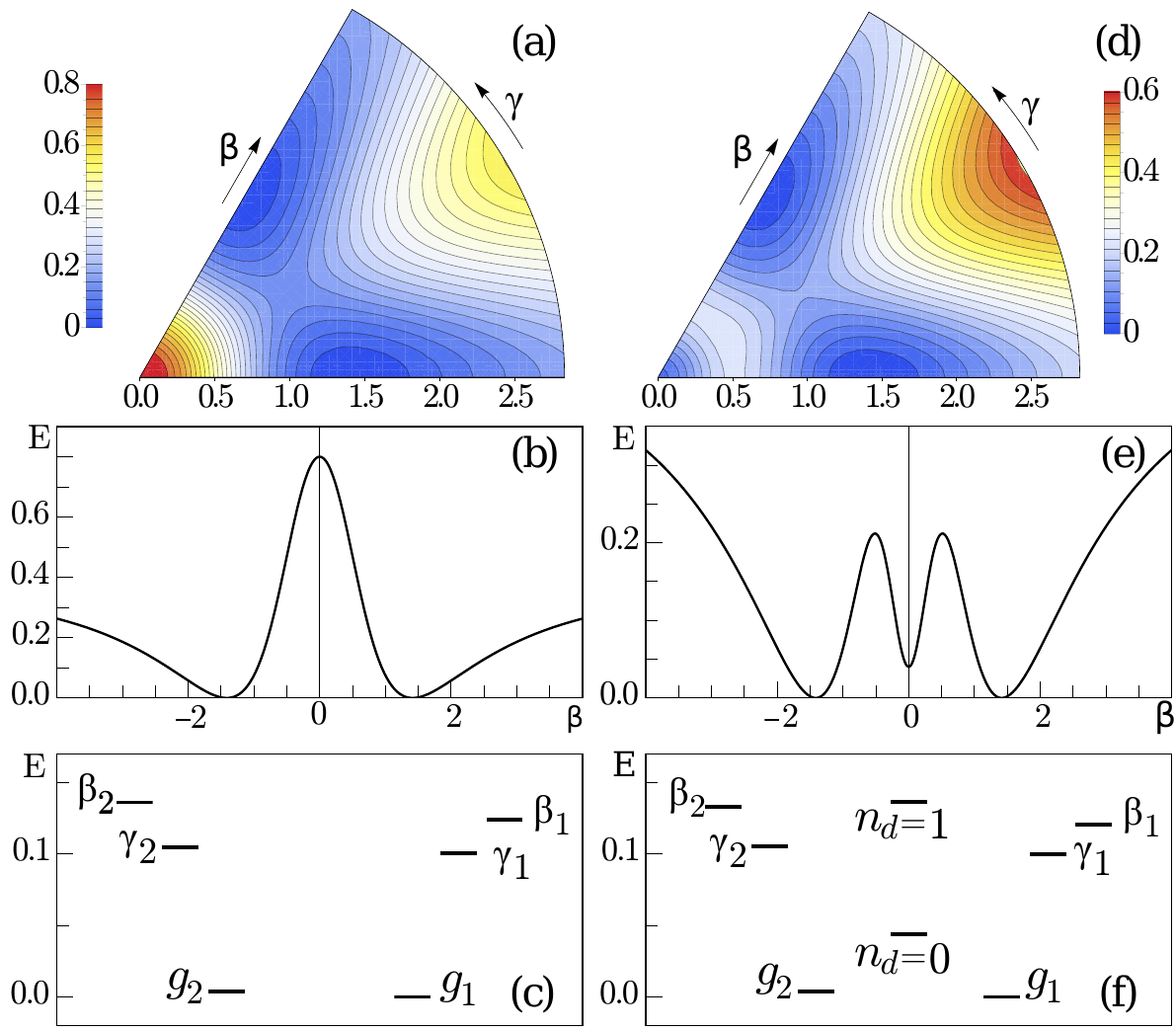}
  \caption{
Contour plots of the energy surface~(\ref{surface}) [top row], 
$\gamma\!=\!0$ sections [middle row] and bandhead spectrum [bottom row] 
for the Hamiltonian~(\ref{Hprime}), with 
$\alpha\!=\!0.018,\,\eta_3=0.571,\,\rho\!=\!0,\,N\!=\!20$. 
Panels (a)-(b)-(c) [(d)-(e)-(f)] correspond to the choice 
$h_0\!=\!0.2,\,h_2\!=\!0.4$ [$h_0\!=\!0.01,\,h_2\!=\!0.5$] resulting in 
prolate-oblate [spherical-prolate-oblate] shape coexistence.
\label{fig1}}
\end{figure}

The intrinsic part of the critical-point Hamiltonian is required to 
satisfy
\bsub
\ba
\hat{H}\ket{N,\, (\lambda,\mu)=(2N,0),\,K=0,\, L} &=& 0 ~,
\label{2N0}
\\
\hat{H}\ket{N,\, (\blam,\bmu)=(0,2N),\,\bar{K}=0,\, L} &=& 0 ~.
\label{02N}
\ea
\label{vanish}
\esub 
Equivalently, $\hat{H}$ annihilates the intrinsic states of 
Eq.~(\ref{int-state}), with $(\beta\!=\!\sqrt{2},\gamma\!=\!0)$ and 
$(\beta\!=\!-\sqrt{2},\gamma\!=\!0)$, which are the lowest and 
highest-weight vectors in the irreps $(2N,0)$ and $(0,2N)$ 
of SU(3) and $\bsu3$, respectively. 
The resulting Hamiltonian is found to be
\ba
\hat{H} = 
h_0\,P^{\dag}_0\hat{n}_sP_0 + h_2\,P^{\dag}_0\hat{n}_dP_0 
+\eta_3\,G^{\dag}_3\cdot\tilde{G}_3 ~,
\label{Hint}
\ea
where $P^{\dag}_{0} = d^{\dagger}\cdot d^{\dagger} - 2(s^{\dagger})^2$, 
$G^{\dag}_{3,\mu} = \sqrt{7}[(d^{\dag} d^{\dag})^{(2)}d^{\dag}]^{(3)}_{\mu}$, 
$\tilde{G}_{3,\mu} = (-1)^{\mu}G_{3,-\mu}$, 
$\hat{n}_d=\sum_{\mu}d^{\dag}_{\mu}d_{\mu}$ 
and the centered dot denotes a scalar product. 
The corresponding energy surface, 
$E_{N}(\beta,\gamma) = N(N-1)(N-2)\tilde{E}(\beta,\gamma)$, 
is given by
\ba
\tilde{E}(\beta,\gamma) &=& 
\left\{(\beta^2-2)^2
\left [h_0 + h_2\beta^2\right ] 
+\eta_3 \beta^6\sin^2(3\gamma)\right \}\quad
\nonumber\\
&&
\times
(1+\beta^2)^{-3} ~.
\label{surface}
\ea
The surface is an even function of $\beta$ and 
$\Gamma = \cos 3\gamma$, 
and can be transcribed as 
$\tilde{E}(\beta,\gamma) = z_0 + 
(1+\beta^2)^{-3}[A\beta^6+ B\beta^6\Gamma^2 + D\beta^4+ F\beta^2]$, 
with $A \!=\! -4h_0 \!+\!h_2 \!+\! \eta_3,\, B \!=\! -\eta_3, \,
D \!=\! -(11h_0 \!+\! 4h_2), \; F \!=\! 4(h_2\!-\!4h_0),\,z_0 \!=\! 4h_0$. 
It is the most general form 
of a surface accommodating 
degenerate prolate and oblate extrema
with equal $\beta$-deformations, 
for an Hamiltonian with cubic terms~\cite{isacker81,levshao90}. 
For $h_0,h_2,\eta_3\geq 0$, 
$\hat{H}$ is positive definite and 
$\tilde{E}(\beta,\gamma)$ has two degenerate global minima, 
$(\beta=\sqrt{2},\gamma=0)$ and $(\beta=\sqrt{2},\gamma=\pi/3)$ 
[or equivalently $(\beta=-\sqrt{2},\gamma=0)$], at $\tilde{E}=0$.
$\beta=0$ is always an extremum, which is a local minimum (maximum) for 
$F>0$ ($F<0$), at $\tilde{E}=4h_0$.
Additional extremal points include 
(i)~a saddle point: $[\bs^2 = \frac{2(4h_0-h_2)}{h_0 - 7h_2}, 
\gamma=0,\pi/3]$, at $\tilde{E}=\frac{4(h_0+2h_2)^3}{81(h_0-h_2)^2}$.
(ii)~A~local maximum and a saddle point: $[\bss^2,\gamma=\pi/6]$, 
at $\tilde{E}= \frac{1}{3}(1+\bss^2)^{-2}\bss^2[D\bss^2+2F] +z_0$, 
where $\bss^2$ is a solution of 
$(D-3A)\bss^4 + 2(F-D)\bss^2 -F = 0$. 
The saddle points, when they exist, support 
a barrier separating the various minima, as seen in Fig.~\ref{fig1}.

The members of the prolate and oblate ground-bands, 
Eq.~(\ref{vanish}), 
are zero-energy eigenstates of $\hat{H}$ (\ref{Hint}), 
with good SU(3) and $\bsu3$ symmetry, 
respectively. For large $N$, the spectrum 
is harmonic, involving $\beta$ and $\gamma$ vibrations about the 
respective deformed minima, with frequencies
\bsub
\ba
&&
\epsilon_{\beta 1}=\epsilon_{\beta 2} 
= \frac{8}{3}(h_0+ 2h_2)N^2 ~,
\\
&&\epsilon_{\gamma 1}=\epsilon_{\gamma 2} = 4\eta_3N^2 ~.
\ea
\label{d-modes}
\esub
For $h_0=0$, also $\beta=0$ becomes a global minimum at $\tilde{E}=0$, 
resulting in three degenerate minima, corresponding to 
triple coexistence of prolate, oblate and 
spherical shapes. $\hat{H}(h_0=0)$ has the following U(5) basis state 
\ba
\hat{H}(h_0=0)\ket{N,n_d=\tau=L=0} = 0 ~,
\ea
as an eigenstate. Equivalently, it annihilates the intrinsic state 
of Eq.~(\ref{int-state}), with $\beta=0$. The additional normal modes 
involve quadrupole vibrations about the spherical minimum, with frequency
\ba
\epsilon = 4h_2N^2 ~.
\label{s-modes}
\ea
When $\beta=0$ is only a local minimum, the $(n_d\!=\!L\!=\!0)$ 
state experiences a shift of order $4h_0N^3$.
\begin{figure}[t]
  \centering \includegraphics[width=8.6cm]{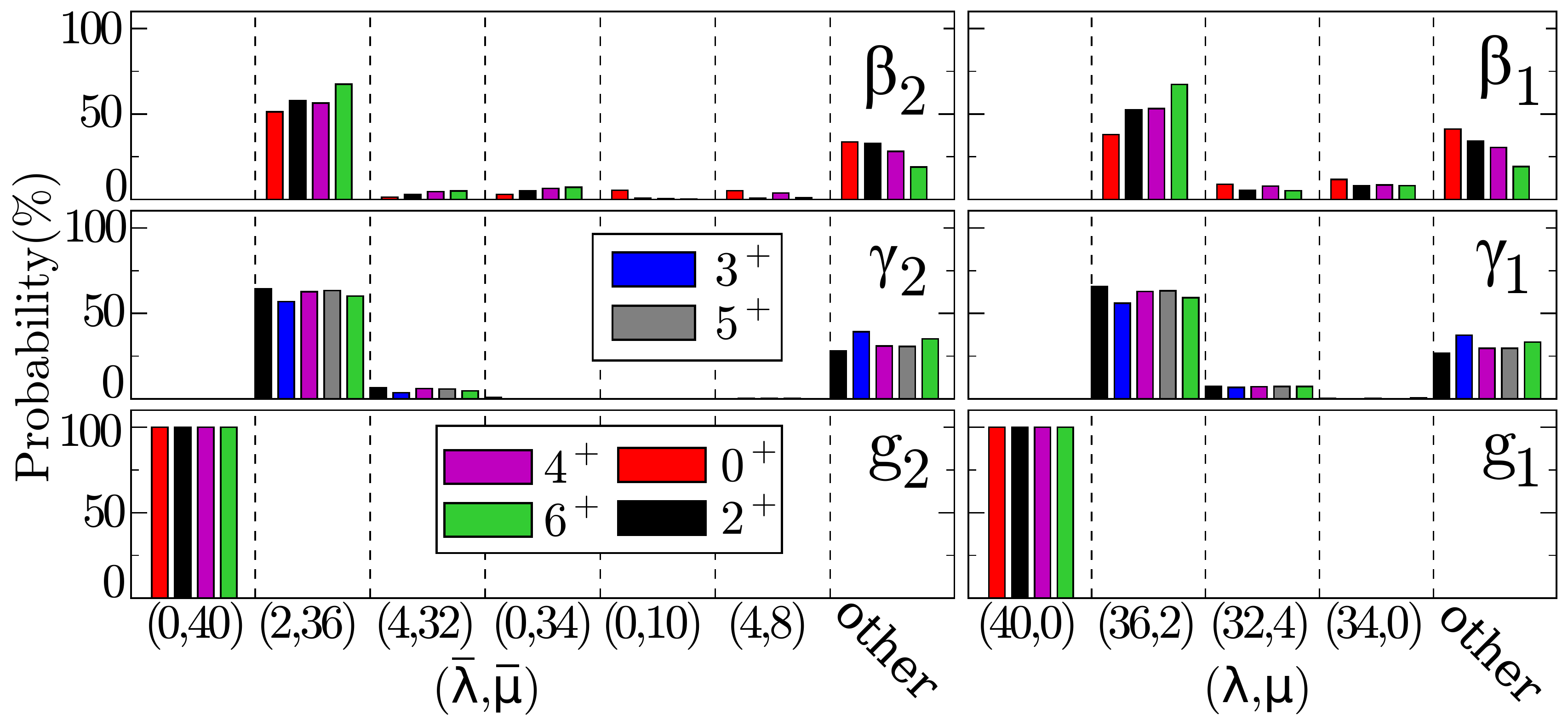}
  \caption{
SU(3) $(\lambda,\mu)$-
and $\bsu3$ $(\blam,\bmu)$- decomposition 
for members of the prolate ($g_1,\beta_1,\gamma_1$) 
and oblate ($g_2,\beta_2,\gamma_2$) bands, eigenstates of 
$\hat{H}'$ (\ref{Hprime}) with parameters as in Fig.~\ref{fig1}(c), 
resulting in prolate-oblate shape coexistence.
The column `other' depicts a sum of probabilities, each less than 0.5\%. 
\label{fig2} }
\end{figure}

The Hamiltonian of Eq.~(\ref{Hint}) is invariant under a change of sign 
of the $s$-bosons, hence commutes with the ${\cal R}_{s}$ operator 
mentioned above. Consequently, 
all non-degenerate eigenstates of $\hat{H}$ 
have well-defined $s$-parity. 
This implies vanishing quadrupole moments for an $E2$ operator 
which is odd under such sign change.
To overcome this difficulty, we introduce a small $s$-parity 
breaking term, ${\textstyle\alpha\hat{\theta}_2 = 
\alpha[-\hat{C}_{2}[SU(3)] + 2\hat{N}(2\hat{N}+3)]}$, 
which contributes to $\tilde{E}(\beta,\gamma)$, 
${\textstyle\tilde{\alpha}(1+\beta^2)^{-2}[ 
(\beta^2\!-\!2)^2 \!+\! 2\beta^2(2 \!-\!2\sqrt{2}\beta\Gamma 
\!+\!\beta^2)]}$, 
with ${\textstyle\tilde{\alpha}=\alpha/(N-2)}$. 
The linear $\Gamma$-dependence distinguishes 
the two deformed minima and slightly lifts 
their degeneracy, as well as that of the normal modes~(\ref{d-modes}). 
Replacing $\hat{\theta}_2$ by 
${\textstyle\bar{\theta}_2}$, associated 
with $\hat{C}_{2}[\bsu3]$, leads to similar effects but
interchanges the role of prolate and oblate bands. 
Identifying the collective part with $\hat{C}_2[{\rm SO(3)}]$, 
we arrive at the following complete Hamiltonian 
\ba
\hat{H}' &=& \hat{H}(h_0,h_2,\eta_3) 
+ \alpha\,\hat{\theta}_2 
+ \rho\,\hat{C}_2[\rm SO(3)] ~,
\label{Hprime}
\ea 
where $\hat{H}(h_0,h_2,\eta_3)$ is the Hamiltonian of Eq.~(\ref{Hint}).
\begin{figure}[t]
  \centering \includegraphics[width=8.6cm]{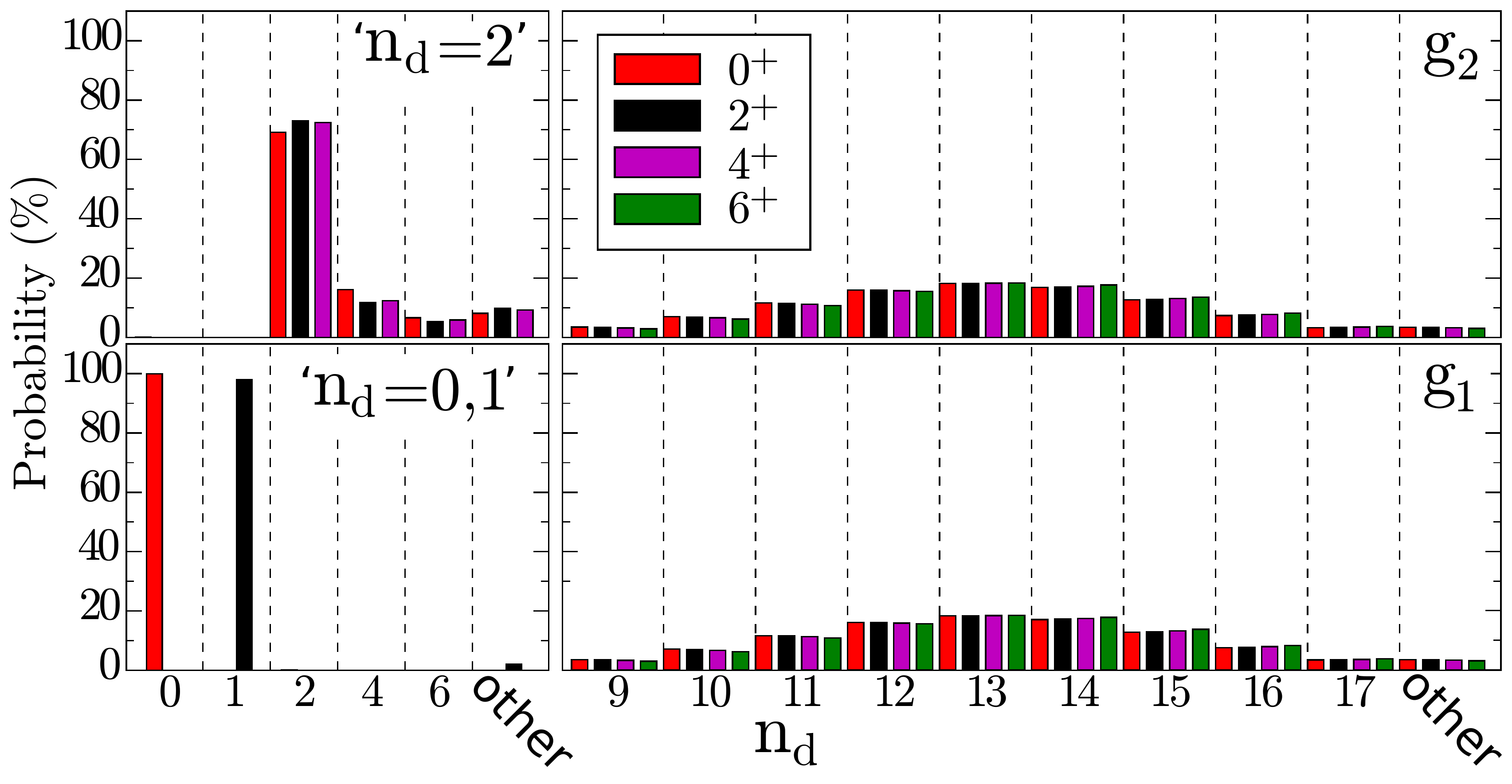}
  \caption{
U(5) $n_d$- decomposition for spherical states (left panels)
and for members of the $g_1$ and $g_2$ bands (right panels), 
eigenstates of 
$\hat{H}'$ (\ref{Hprime}) with parameters as in Fig.~\ref{fig1}(f), 
resulting in spherical-prolate-oblate shape coexistence.
\label{fig3}}
\end{figure}

Figures 1(a)-1(b)-1(c) [1(d)-1(e)-1(f)] of Fig.~\ref{fig1}, show 
$\tilde{E}(\beta,\gamma)$, $\tilde{E}(\beta,\gamma\!=\!0)$ 
and the bandhead spectrum of $\hat{H}'$ (\ref{Hprime}), 
with parameters ensuring P-O [S-P-O] minima. The prolate $g_1$-band 
remains solvable with energy $E_{g1}(L)=\rho L(L+1)$. 
The oblate $g_2$-band experiences a slight shift of 
order ${\textstyle\tfrac{32}{9}\alpha N^2}$ and 
displays a rigid-rotor like spectrum. In the case of P-O coexistence, 
the SU(3) and $\bsu3$ decomposition in Fig.~\ref{fig2} demonstrates 
that these bands are pure DS basis states, with 
$(2N,0)$ and $(0,2N)$ character, respectively, 
while excited $\beta$ and $\gamma$ bands exhibit considerable mixing. 
In the case of triple S-P-O coexistence, the prolate and oblate bands show 
similar behaviour. A new aspect is the 
simultaneous occurrence in the spectrum [Fig.~\ref{fig1}(f)] 
of spherical type of states, whose wave functions are dominated by 
a single $n_d$ component. 
As shown in Fig.~\ref{fig3}, 
the lowest spherical states have quantum numbers 
$(n_d\!=\!L\!=\!0)$ and $(n_d\!=\!1,L\!=\!2)$, 
hence coincide with pure U(5) basis states, while higher spherical states 
have a pronounced ($\sim$70\%) $n_d\!=\!2$ component. This structure 
should be contrasted with the U(5) decomposition of deformed states 
(belonging to the $g_1$ and $g_2$ bands) which, as shown in 
Fig.~\ref{fig3}, have a broad $n_d$-distribution.
The purity of selected sets of states 
with respect to SU(3), $\bsu3$ and U(5), in the presence of other 
mixed states, are the hallmarks of a partial 
dynamical symmetry~\cite{Leviatan96,Leviatan11}. 
It is remarkable that a simple Hamiltonian, as in Eq.~(\ref{Hprime}), 
can accommodate simultaneously several incompatible symmetries in 
a segment of the spectrum.

Since the wave functions for the members of the $g_1$ and $g_2$ bands 
are known, one can derive analytic expressions for their 
quadrupole moments and $E2$ transition rates. 
Considering the $E2$ operator
\ba
T(E2) = e_B(d^{\dag}s+s^{\dag}\tilde{d}) ~, 
\label{Te2}
\ea
with an effective charge $e_B$, 
the quadrupole moments 
are found to have equal magnitudes and opposite signs, 
\ba
Q_L &=& 
{\textstyle
\mp e_B\sqrt{\frac{16\pi}{40}}\frac{L}{2L+3}
\frac{4(2N-L)(2N+L+1)}{3(2N-1)}} ~,
\label{quadmom}
\ea
where the minus (plus) sign corresponds to the prolate-$g_1$ (oblate-$g_2$) 
band. The B($E2$) values for intra-band ($g_1\to g_1$, $g_2\to g_2$) 
transitions, 
\ba
&&B(E2; g_i,\, L+2\to g_i,\,L) = 
\nonumber\\
&&
\quad
\;\;
{\textstyle
e_{B}^2\,\frac{3(L+1)(L+2)}{2(2L+3)(2L+5)}
\frac{(4N-1)^2(2N-L)(2N+L+3)}{18(2N-1)^2}} ~,
\qquad\qquad
\label{be2}
\ea
are the same. These properties are ensured by the fact that 
${\cal R}_sT(E2){\cal R}_s^{-1} = -T(E2)$. Inter-band 
$(g_2\leftrightarrow g_1)$ 
$E2$ transitions, are extremely weak. This follows from the fact that 
the $L$-states of the $g_1$ and $g_2$ bands exhaust, respectively, 
the $(2N,0)$ and $(0,2N)$ irrep of SU(3) and $\bsu3$. 
T($E2$) as a $(2,2)$ tensor under both algebras can 
thus connect the $(2N,0)$ irrep of $g_1$ 
only with the $(2N-4,2)$ component in $g_2$ which, however, is 
vanishingly small. The selection rule $g_1\nleftrightarrow g_2$ 
is valid also for a more general $E2$ operator, 
obtained by adding $Q^{(2)}$ or $\bar{Q}^{(2)}$ to 
the operator of Eq.~(\ref{Te2}) 
since the latter, as generators, 
cannot mix different irreps of SU(3) or $\bsu3$. 
By similar arguments, $E0$ transitions in-between 
the $g_1$ and $g_2$ bands are extremely weak, 
since the relevant operator, 
$T(E0)\propto\hat{n}_d$, is a combination of $(0,0)$ and $(2,2)$ 
tensors under both algebras. Accordingly, 
the $L=0$ bandhead state of the higher ($g_2$) band, 
cannot decay to the lower $g_1$ band, hence displays characteristic 
features of an isomeric state. In contrast to $g_1$ and $g_2$, excited 
$\beta$ and $\gamma$ bands are mixed, hence are connected by $E2$ transitions 
to these ground bands. 
Their quadrupole moments are found numerically to resemble, for large $N$, 
the collective model expression 
$Q(K,L) = \frac{3K^2-L(L+1)}{(L+1)(2L+3)} q_{K}$,
with $q_{K}>0$ ($q_{K}<0$) for prolate (oblate) bands.

In the case of triple (S-P-O) coexistence, since $T(E2)$ 
obeys the selection rule $\Delta n_d\!=\!\pm 1$, the 
spherical states, $(n_d\!=\!L\!=\!0)$ and $(n_d\!=\!1,L\!=\!2)$,
have no quadrupole moment and the B($E2$) value for their 
connecting transition, obeys the U(5)-DS expression~\cite{ibm}
\ba
B(E2; n_d=1,L=2\to n_d=0,L=0) = e_{B}^2N ~.\quad
\label{be2nd}
\ea
These spherical states have very weak $E2$ transitions to the 
deformed ground bands, because they exhaust the $(n_d\!=\!0,1)$ irreps 
of U(5), and the $n_d\!=\!2$ component in the ($L\!=\!0,2,4$) states 
of the $g_1$ and $g_2$ bands is 
extremely small, of order $N^33^{-N}$. 
There are also no $E0$ transitions involving these spherical states, 
since $T(E0)$ is diagonal in $n_d$.
The lowest ($n_d\!=\!L\!=\!0$) state has, therefore, 
the attributes of a spherical isomer state.
The analytic expressions 
of Eqs.~(\ref{quadmom})-(\ref{be2nd}) are parameter-free predictions, 
except for a scale, and can be used 
to compare with measured values of these observables 
and to test the underlying SU(3), $\bsu3$ and U(5) partial symmetries.

The proposed Hamiltonian~(\ref{Hprime}) involves three-body interactions. 
Similar cubic terms were encountered in previous 
studies within the IBM, in conjunction with 
triaxiality~\cite{heyde84,zamfir91}, 
band anharmonicity~\cite{Ramos00,GarciaRamos09} 
and signature splitting~\cite{Bonatsos88,Leviatan13} in deformed nuclei. 
Higher-order terms show up naturally in microscopic-inspired IBM Hamiltonians
derived by a mapping from self-consistent mean-field 
calculations~\cite{nomura13,Nomura12}. 
Near shell-closure $\hat{H}'$~(\ref{Hprime}) 
can be regarded as an effective number-conserving 
Hamiltonian, which simulates the excluded 
intruder-configurations by means of higher-order terms. Indeed, 
the energy surfaces of the IBM with configuration 
mixing~\cite{Frank04,Morales08,hellemans09} 
contain higher-powers of $\beta^2$ 
and $\beta^3\cos3\gamma$, as in Eq.~(\ref{surface}). 
Recalling the microscopic interpretation of the IBM bosons 
as images of identical valence-nucleon pairs, the results of the present study 
suggest that for nuclei far from shell-closure, shape coexistence can occur 
within the same valence space.

As discussed, 
the coexisting prolate and oblate ground bands of $\hat{H}'$~(\ref{Hprime}), 
are unmixed and retain their individual SU(3) and $\bsu3$ character. 
This situation is different from that encountered in the neutron-deficient 
Kr~\cite{Clement07} and Hg~\cite{Bree14} isotopes, 
where the observed structures are strongly mixed. 
It is in line with the recent evidence for shape-coexistence 
in neutron-rich Sr isotopes, where 
spherical and prolate-deformed configurations exhibit very weak 
mixing~\cite{Clement16}. Band mixing can be incorporated in the present 
formalism by adding kinetic rotational terms which do 
not affect the shape of the energy 
surface~\cite{kirlev85,Leviatan87,levkir90,levshao90}, 
but may destroy the partial symmetry property of the states. 
The evolution of structure away from the critical point, can be studied by 
varying the coupling constant $\alpha$ in Eq.~(\ref{Hprime}). 
Larger values of $\alpha$ will shift the energy of the 
non-yrast ground band (the oblate $g_2$ band in the example considered). 
In the case of a triple S-P-O coexistence, adding an $\hat{n}_d$ term 
to $\hat{H}(\eta_0=0)$, will leave the $n_d=0$ spherical ground state 
unchanged, but will shift the prolate and oblate bands to higher energy.
The same method of intrinsic-collective resolution 
can be used to identify appropriate Hamiltonians for an 
asymmetric prolate-oblate coexistence with different $\beta$-deformations. 
Details of such extensions and refinements will be reported elsewhere.

In summary, we have presented a number-conserving rotational-invariant 
Hamiltonian which captures essential features of P-O and S-P-O 
coexistence in nuclei. 
It preserves particular symmetries for certain prolate and oblate bands 
and spherical states, with closed expressions for $E2$ moments and 
transition rates, which are the observables most closely related to the 
nuclear shape. These attributes turn the proposed framework into 
a suitable algebraic benchmark for the study of shape-coexistence 
in nuclei, providing a convenient starting point, guidance and test-ground 
for more detailed treatments of this intriguing phenomena. 
This research was supported by the Israel Science Foundation 
(Grant No. 493/12).


\begin{thebibliography}{[99]}

\bibitem{Heyde11}
K. Heyde and J.L. Wood, 
Rev. Mod. Phys. {\bf 83}, 1467 (2011).

\bibitem{Focus16}
Edited by J.L. Wood and K. Heyde, 
{\it Focus Review on Shape Coexistence in Nuclei}, 
J. Phys. G {\bf 43}, 024001-024013 (2016).

\bibitem{Jenkins14}
D.G. Jenkins, 
Nature Phys. {\bf 10}, 909 (2014).

\bibitem{Clement07}
E. Cl\'ement {\it et al.},
Phys. Rev. C {\bf 75}, 054313 (2007).

\bibitem{Ljun08}
J. Ljungvall {\it et al.},
Phys. Rev. Lett. {\bf 100}, 102502 (2008).

\bibitem{Bree14}
N. Bree et al.
Phys. Rev. Lett. {\bf 112}, 162701 (2014).

\bibitem{Andreyev00}
A.N. Andreyev {\it et al.}, 
Nature {\bf 405}, 430 (2000).

\bibitem{MCSM99}
T. Mizusaki, T. Otsuka, Y. Utsuno, M. Honma, and T. Sebe, 
Phys. Rev. C {\bf 59}, 1846(R) (1999).

\bibitem{Tsunuda14}
Y. Tsunoda, T. Otsuka, N. Shimizu, M. Honma, and Y. Utsuno,
Phys. Rev. C {\bf 89}, 031301(R) (2014).

\bibitem{Foisson03}
R. Fossion, K. Heyde, G. Thiamova and P. Van Isacker
Phys. Rev. C {\bf 67}, 024306 (2003).

\bibitem{Frank04}
A. Frank, P. Van Isacker and C. E. Vargas, 
Phys. Rev. C {\bf 69}, 034323 (2004).

\bibitem{Morales08}
I. O. Morales, A. Frank, C.E. Vargas and P. Van Isacker,
Phys. Rev. C {\bf 78}, 024303 (2008).

\bibitem{ramos14}
J.~E.~Garc\'\i a-Ramos and K. Heyde,
Phys. Rev. C {\bf 89}, 014306 (2014).

\bibitem{robeldo12}
K. Nomura, R. Rodr\'\i guez-Guzm\' an, L. M. Robledo, and N. Shimizu,
Phys. Rev. C {\bf 86}, 034322 (2012).

\bibitem{nomura13}
K. Nomura, R. Rodr\'\i guez-Guzm\' an and L. M. Robledo,
Phys. Rev. C {\bf 87}, 064313 (2013).

\bibitem{Nomura16}
K. Nomura, T. Otsuka and P. Van Isacker, 
J. Phys. G {\bf 43}, 024008 (2016).

\bibitem{Bender13}
J. M. Yao, M. Bender and P.-H. Heenen,
Phys. Rev. C {\bf 87}, 034322 (2013).

\bibitem{vretenar16}
Z.P. Li, T. Nik\v si\'c and D. Vretenar,
J. Phys. G {\bf 43}, 024005 (2016).

\bibitem{EDF-IBM11}
K.~Nomura, T.~Nik\v si\'c, T.~Otsuka, N.~Shimizu and D. Vretenar, 
Phys. Rev. C {\bf 84}, 014302 (2011).

\bibitem{moller09}
P. M\"oller, A.J. Sierk, R. Bengtsson, H. Sagawa and T. Ichikawa,
Phys. Rev. Lett. {\bf 103}, 212501 (2009).

\bibitem{moller12}
P. M\"oller, A.J. Sierk, R. Bengtsson, H. Sagawa and T. Ichikawa,
Atomic Dat. Nucl. Dat. Tables {\bf 98}, 149 (2012).

\bibitem{ibm}
F.~Iachello and A.~Arima,
{\it The Interacting Boson Model}
(Cambridge University Press, Cambridge, 1987).

\bibitem{gino80}
J.N. Ginocchio and M.W. Kirson, 
Phys. Rev. Lett. {\bf 44}, 1744 (1980).

\bibitem{diep80}
A.E.L. Dieperink, O. Scholten and F. Iachello,
Phys. Rev. Lett. {\bf 44}, 1747 (1980).

\bibitem{jolie01}
J. Jolie, R.F. Casten, P. von Brentano and V. Werner, 
Phys. Rev. Lett. {\bf 87}, 162501 (2001).

\bibitem{jolie03}
J. Jolie and A. Linnemann, 
Phys. Rev. C {\bf 68}, 031301(R) (2003).

\bibitem{draayer12}
Y. Zhang, F. Pan, Y.X. Liu, Y.A. Luo and J. P. Draayer,
Phys. Rev. C {\bf 85}, 064312 (2012).

\bibitem{kirlev85}
M.W. Kirson and A. Leviatan,
Phys. Rev. Lett. {\bf 55}, 2846 (1985). 

\bibitem{Leviatan87}
A. Leviatan, 
Ann. Phys. (N.Y.) {\bf 179}, 201 (1987).

\bibitem{levkir90}
A. Leviatan and M.W. Kirson,
Ann. Phys. (N.Y.) {\bf 201}, 13 (1990).

\bibitem{Leviatan06}
A. Leviatan,
Phys. Rev. C {\bf 74}, 051301(R)  (2006). 

\bibitem{Leviatan07}
A.~Leviatan,
Phys. Rev. Lett. {\bf 98}, 242502 (2007).

\bibitem{Macek14}
M.~Macek and A.~Leviatan,
Ann. Phys. (N.Y.) {\bf 351}, 302 (2014).

\bibitem{isacker81}
P. Van Isacker and J.Q. Chen,
Phys. Rev. C {\bf 24}, 684 (1981).

\bibitem{levshao90}
A. Leviatan and B. Shao, 
Phys. Lett. B {\bf 243}, 313 (1990).

\bibitem{Leviatan96}
A.~Leviatan, 
Phys. Rev. Lett. {\bf 77}, 818 (1996).

\bibitem{Leviatan11}
A.~Leviatan, 
Prog. Part. Nucl. Phys. {\bf 66}, 93 (2011).

\bibitem{heyde84}
K. Heyde, P. Van Isacker, M. Waroquier, and J. Moreau,
Phys. Rev. C {\bf 29}, 1420 (1984).

\bibitem{zamfir91}
N.V. Zamfir and R.F. Casten,
Phys. Lett. B {\bf 260}, 265 (1991).

\bibitem{Ramos00}
J.~E.~Garc\'\i a-Ramos, J.M. Arias and P.~Van~Isacker,
Phys. Rev. C {\bf 62}, 064309 (2000).

\bibitem{GarciaRamos09}
J.~E.~Garc\'\i a-Ramos, A.~Leviatan, and P.~Van~Isacker,
Phys. Rev. Lett. {\bf 102}, 112502 (2009).

\bibitem{Bonatsos88}
D. Bonatsos, 
Phys. Lett. B {\bf 200}, 1 (1988).

\bibitem{Leviatan13}
A.~Leviatan, J.~E.~Garc\'\i a-Ramos, and P.~Van~Isacker,
Phys. Rev. C {\bf 87}, 021302(R) (2013).

\bibitem{Nomura12}
K.~Nomura, N.~Shimizu, D.~Vretenar, T.~Nik\v si\'c and T.~Otsuka,
Phys. Rev. Lett. {\bf 108}, 132501 (2012).

\bibitem{hellemans09}
V. Hellemans, P. Van Isacker, S. De Baerdemacker and K. Heyde, 
Nucl. Phys. A {\bf 819},  11 (2009). 

\bibitem{Clement16}
E. Cl\'ement {\it et al.}, 
Phys. Rev. Lett. {\bf 116}, 022701 (2016).

\end{thebibliography}
\end{document}